\documentclass[%
reprint,
amsmath,amssymb,
aps,
pra
]{revtex4-2}

\usepackage{graphicx}
\usepackage{dcolumn}
\usepackage{bm}

\begin{document}
\preprint{APS/123-QED}
	
\title{Measuring orbital angular momentum of vortex beams in optomechanics} 
\author{Zhucheng Zhang,$^{1}$ Jiancheng Pei,$^{1}$ Yi-Ping Wang,$^{2}$ and Xiaoguang Wang$^{1,3}$}
\email{xgwang1208@zju.edu.cn}
\affiliation{$^{1}$Zhejiang Institute of Modern Physics, Department of Physics, Zhejiang University, Hangzhou 310027, China\\
$^{2}$College of Science, Northwest A$\&$F University, Yangling 712100, China\\
$^{3}$Graduate School of China Academy of Engineering Physics, Beijing 100193, China}
\date{\today}

\begin{abstract}
Measuring the orbital angular momentum (OAM) of vortex beams, including the magnitude and the sign,
has great application prospects due to its theoretically unbounded and orthogonal modes.
Here, the sign-distinguishable OAM measurement in optomechanics
is proposed, which is achieved by monitoring the shift of the transmission spectrum of the probe field in a double
Laguerre-Gaussian (LG) rotational-cavity system. Compared with the traditional single LG rotational cavity,
an asymmetric optomechanically induced transparency window can occur in our system.
Meanwhile, the position of the resonance valley has a strong correlation with the magnitude and sign of OAM.
This originally comes from the fact that the effective detuning of the cavity mode from the driving field
can vary with the magnitude and sign of OAM, which causes the spectral shift to be directional for different signs of OAM.
Our scheme solves the shortcoming of the inability to distinguish the sign of OAM in optomechanics,
and works well for high-order vortex beams with topological charge value $\pm 45$,
which is a significant improvement for measuring OAM based on the cavity optomechanical system.
\begin{description}
	\item[keywords]{orbital angular momentum, optomechanically induced transparency, Laguerre-Gaussian rotational-cavity system, optomechanics}
\end{description}
\end{abstract}

\maketitle

\section{Introduction}
Vortex beams, such as Laguerre-Gaussian (LG) beam, possess an azimuthal phase structure $e^{il\psi}$, which can
carry a well-defined orbital angular momentum (OAM) of $l \hbar$ per photon, with $\psi$ and $l$ being its azimuthal
angle and topological charge value \cite{nye1974dislocations,allen1992orbital}. This type of beams can be generated by
diffracting a non-helical beam off a spiral phase plate \cite{beijersbergen1994helical-wavefront,oemrawsingh2004half-integral} or 
off a computer-generated hologram \cite{bazhenov1992screw,basistiy1993optics}.
Recently, the generation and detection of OAM-tunable vortex microlaser on the photonic chip were realized \cite{Zhang2020Tunable,Ji2020Photocurrent}.
Due to their quantized OAM and their dynamic characteristics, these
helically phased beams are widely used in many fields, 
such as quantum information technologies \cite{ding2015Quantum}, optical communications \cite{wang2012terabit,bozinovic2013terabit-scale}, optical trapping \cite{chen2013dynamics}, 
optical tweezers \cite{padgett2011tweezers,gecevicius2014single}, and so on. Thus, it is of great importance to measure OAM of vortex
beams (or its topological charge value) accurately, including the magnitude and the sign.

To measure OAM of vortex beams, in general, we can analyze the related interference patterns directly, for example,
the interference pattern between the spiral wave front and a flat wave front \cite{harris1994optical}, or the interference pattern between a vortex
beam and its mirror image \cite{harris1994laser}. With the use of a triangular aperture and an annular aperture, the measurements of the
topological charge value with $l=\pm 7$ and $|l|=9$ based on the diffraction pattern were also reported, respectively \cite{hickmann2010unveiling,guo2009characterizing}.
Besides, the measurable value of the topological charge was raised to $\pm 14$ and $\pm25$ by using a tilted convex lens \cite{vaity2013measuring} and annular gratings \cite{zheng2017measuring}. Recently,
based on the optomechanically induced transparency (OMIT) phenomenon \cite{harris1997electromagnetically,agarwal2010electromagnetically,weis2010optomechanically}
generated in a single LG rotational-cavity system, the topological charge value ranging from $0$ to $42$ can be measured in theory
\cite{peng2019optomechanically}, but this scheme cannot distinguish the sign of OAM.

In this paper, we propose a scheme to measure OAM, with a distinguishable sign and a wider range, in a double LG rotational-cavity optomechanical system.
The LG rotational-cavity system, composed of spiral phase plates \cite{beijersbergen1994helical-wavefront,oemrawsingh2004half-integral}, was first proposed
by Bhattacharya and Meystre to trap and cool the rotational motion of a mirror \cite{bhattacharya2007using}. In
this type of cavity optomechanical system, the intracavity radiation field can exchange linear and angular momentum
with the mirror, which is the difference from the traditional cavity optomechanical system \cite{Aspelmeyer2014Cavity,law1995interaction,bhattacharya2008optomechanical,xiao2014controllable,zhang2020photon-assisted}.  And later, many interesting
physical effects have been also studied, such as the entanglement phenomenon based on the LG rotational-cavity system
\cite{bhattacharya2008entanglement,bhattacharya2008entangling,chen2019entanglement}, the ground-state cooling of the rotational mirror in the
unresolved sideband regime \cite{liu2018ground-state}, and OMIT \cite{peng2019optomechanically,peng2020double}. However,
the traditional single LG rotational-cavity system can only distinguish the magnitude of OAM, but not its sign \cite{peng2019optomechanically}.
Then, the OAM measurement in the LG rotational-cavity optomechanical system, with a distinguishable sign and a wider range,
will be interesting and valuable in the field of quantum sensing.

In our system, we show that the effective detuning of the cavity mode from the driving field can vary with the magnitude and sign of OAM simultaneously,
which is different from that of the traditional single LG rotational-cavity system \cite{bhattacharya2007using,peng2019optomechanically}.
In the single LG rotational-cavity system, the effective cavity detuning is only related to the magnitude of OAM but not to its sign, so with the shift of the OMIT
window in Ref.~\cite{peng2019optomechanically}, only the magnitude of OAM can be measured. Compared with the traditional single
LG rotational cavity, we also find that an asymmetric OMIT window can occur in our system. Meanwhile, the position of the resonance valley has a strong correlation with the
magnitude and sign of OAM. By monitoring the position of the resonance valley, the measurable topological charge value can reach to $\pm 45$ in our scheme.
We would like to point out that the measurement of OAM by monitoring the shift of the transmission spectrum of the probe field in our paper can be seen as
the measurement of the \textit{optorotational} coupling between radiation field and rotational mirror \cite{bhattacharya2007using},
which is similar to the \textit{optovibrational} coupling in the traditional optomechanical system with linear momentum exchange. This type of optomechanical coupling was also estimated based on the fisher information
in a recently published work \cite{sanavio2020fisher}. Furthermore, the manufacturing requirements of the spiral phase plate in the LG rotational-cavity optomechanical system are quite severe
\cite{beijersbergen1994helical-wavefront,oemrawsingh2004half-integral}, which may cause the OAM transferred from the rotational mirror to the beams to deviate from the expected value.
Therefore, it is very necessary to measure the OAM carried by the output fields. In short, our scheme solves the shortcoming of the inability to distinguish the sign
of OAM in optomechanics, which is a significant improvement for measuring OAM based on the cavity optomechanical system.

This paper is organized as follows. In Sec.~\ref{2}, we introduce our system model and derive the dynamical equation. In
Sec.~\ref{3}, we discuss in detail the transmission spectrum of the probe field in the double LG rotational cavity, and
compare it with the case of single LG rotational cavity. In Sec.~\ref{4}, we propose our scheme to measure the magnitude
and sign of OAM. Finally, we summarize our conclusions in Sec.~\ref{5}.

\section{Theoretical model}\label{2}
\begin{figure}
	\centering
	\includegraphics[width=1.0\columnwidth]{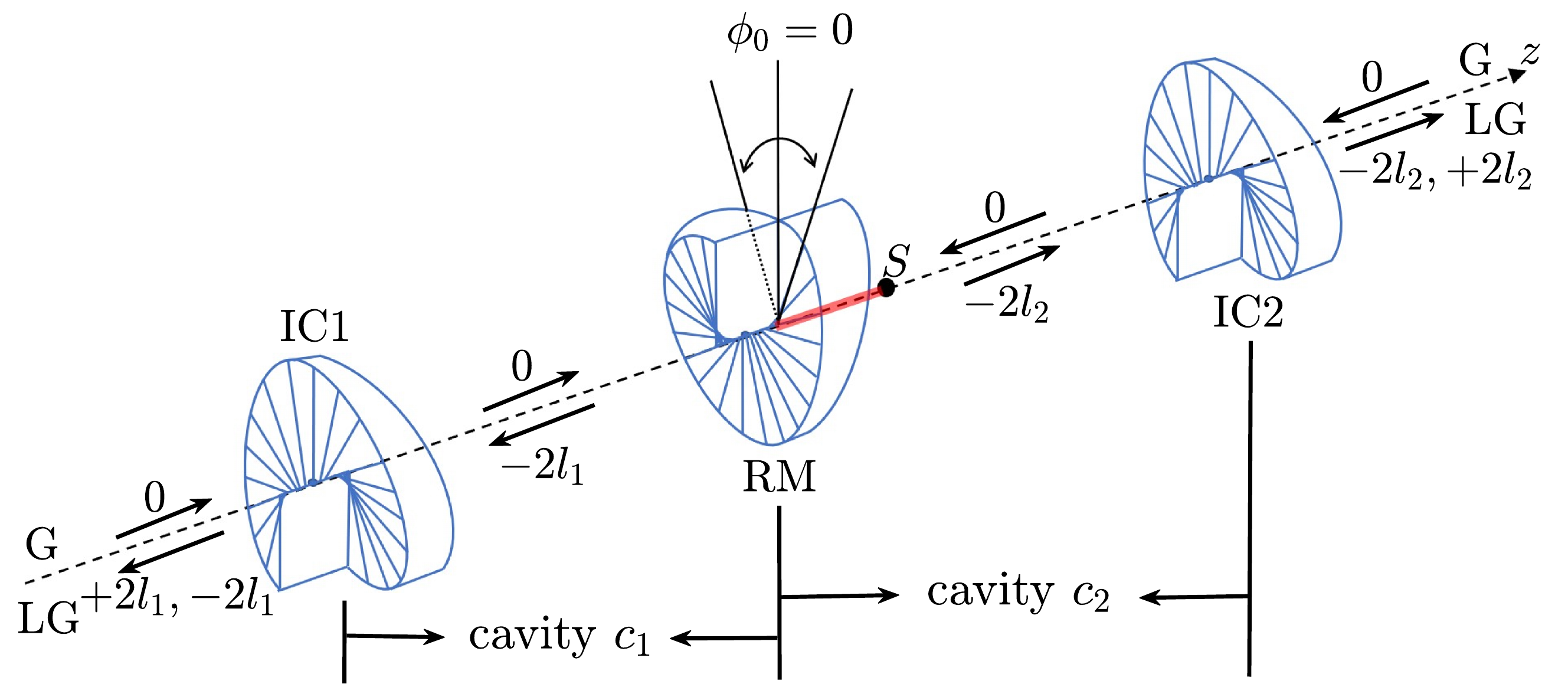}
	\caption{Arrangement for measuring the orbital angular momentum of light in the Laguerre-Gaussian (LG) rotational-cavity system, in which the two input couplers (IC1 and IC2) are partially transparent and rigidly fixed, but the rotational mirror (RM) is perfectly reflective and rotates about the cavity axis $z$ (with angular equilibrium position $\phi_{0}=0$) on a support $S$.  IC1 (IC2) and RM are all spiral phase elements. Input fields injected to the system are Gaussian (G) fields with topological charge 0, and the charge on the LG beams at various points has been indicated. More details about the system can be found in Refs.~\cite{bhattacharya2007using,bhattacharya2008entanglement,bhattacharya2008entangling}.}
	\label{fig:f1}
\end{figure}
In our scheme, we consider a double LG rotational-cavity optomechanical system shown in Fig.~\ref{fig:f1}, which consists of two input
couplers (IC1 and IC2) and a rotational mirror (RM). IC1 (IC2) and RM are all spiral phase elements, which can modify
the azimuthal structure of laser beams \cite{beijersbergen1994helical-wavefront,oemrawsingh2004half-integral}.
RM can be seen as a thin disk with the moment of inertia $I=MR^{2}/2$ ($M$ and $R$ being its mass and radius). We assume
that the input fields injected to the system are all Gaussian (G) fields with topological charge 0, whose charges do
not change when passing through IC1 and IC2 while the reflected components gain a charge $2l_{1}$ and $2l_2$. In addition, 
by changing the step height on both sides of RM or the optical wavelength \cite{oemrawsingh2004half-integral},
RM can remove a topological charge $2l_{1}$ ($2l_2$) from the beams in cavity $c_{1}$ (cavity $c_{2}$). The
related cavity conditions have been discussed in detail in Refs.~\cite{bhattacharya2007using,bhattacharya2008entanglement,bhattacharya2008entangling}.

In our system, the two cavities are assumed to have the same resonance frequency $\omega_c$. Cavity $c_{1}$ is driven by a strong driving field of frequency $\omega_{1}$ and
amplitude $\epsilon_{1}$ and probed by a weak probe field of frequency $\omega_{p}$ and amplitude
$\epsilon_{p}$. Meanwhile, another driving field of frequency $\omega_{2}$ and amplitude
$\epsilon_{2}$ is injected to cavity $c_{2}$. In a rotating frame with respect to the driving fields, the Hamiltonian of
system can be written as
\begin{align}
H&=\hbar\text{(}\Delta_{c1}c_{1}^{\dagger}c_{1}+\Delta_{c2}c_{2}^{\dagger}c_{2})+\frac{L_{z}^{2}}{2I}
+\frac{1}{2}I\omega_{\phi}^{2}\phi^{2} \notag \\
&+\hbar(g_{1}c_{1}^{\dagger}c_{1}-g_{2}c_{2}^{\dagger}c_{2})\phi \notag \\
&+i\hbar[(\epsilon_{1}+\epsilon_{p}e^{-i\Omega t})c_{1}^{\dagger}-\textrm{H.c.}]
+i\hbar(\epsilon_{2}c_{2}^{\dagger}-\textrm{H.c.}),
\end{align}
in which $c_{1}\,(c_{1}^{\dagger})$ and $c_{2}\,(c_{2}^{\dagger})$ are the bosonic annihilation (creation) operators of the
two cavity modes, respectively, satisfying the commutation relation $[c_{1,2},\,c_{1,2}^{\dagger}]=1$.
$\Delta_{c_1,c_2}=\omega_{c}-\omega_{1,2}$ and $\Omega=\omega_{p}-\omega_{1}$ are the
frequency detunings of the driving fields from the cavity modes and the probe field. $L_{z}$ and $\phi$ denote the
angular momentum of RM about the cavity axis $z$ and the angular displacement with the commutation relation
$[L_{z},\,\phi]=-i\hbar$, and $\omega_{\phi}$ is the angular rotation frequency.
$g_{1,2}=cl_{1,2}/L$ characterize the \textit{optorotational} coupling between two LG cavity modes and
RM \cite{bhattacharya2007using,bhattacharya2008entanglement,bhattacharya2008entangling},
with $c$ and $L$ being the speed of light in vacuum and the length of the cavity. The last two terms
describe the coupling between the input fields and the two cavity modes with amplitudes $ \epsilon_{1,2}=\sqrt{2\kappa_{1,2} P_{1,2}/\hbar\omega_{1,2}}$ and $\epsilon_{p}=\sqrt{2\kappa_1 P_{p}/\hbar\omega_{p}}$. $\kappa_{1,2}$
and $P_{1,2,p}$ are the corresponding decay rates of the two cavities and the powers of input fields.

Our scheme focuses on the mean response of the system to the probe field, so we consider the mean-value equations of
the system, which can be obtained by deriving the Heisenberg equations of the system operators as well as adding the
corresponding damping terms. By using the factorization assumption $\left\langle AB\right\rangle =\left\langle A\right\rangle \left\langle B\right\rangle$ \cite{agarwal2010electromagnetically}, the mean value equations of the system operators can be derived as follows,
\begin{align}
\left\langle \frac{dc_{1}}{dt}\right\rangle 	&\!=\!	-[\kappa_1\!+\!i(\Delta_{c1}\!+\!g_{1}\left\langle \phi\right\rangle )]\left\langle c_{1}\right\rangle \!+\!\epsilon_{1}\!+\!\epsilon_{p}e^{-i\Omega t},\\
\left\langle \frac{dc_{2}}{dt}\right\rangle 	&\!=\!	-[\kappa_2+i(\Delta_{c2}-g_{2}\left\langle \phi\right\rangle )]\left\langle c_{2}\right\rangle +\epsilon_{2}, \\
\left\langle \frac{d^{2}\phi}{dt^{2}}\right\rangle 	&\!=\!	
-\gamma_{\phi}\left\langle \frac{d\phi}{dt}\right\rangle -\omega_{\phi}^{2}\left\langle \phi\right\rangle  \notag\\
&-\frac{\hbar}{I}\left(g_{1}\left\langle c_{1}^{\dagger}\right\rangle \left\langle c_{1}\right\rangle -g_{2}\left\langle c_{2}^{\dagger}\right\rangle \left\langle c_{2}\right\rangle \right),
\end{align}
in which $\gamma_{\phi}$ is the damping rate of RM.

The above mean value equations are nonlinear equations, but it can be solved by using the perturbation method due to the
fact that the driving fields are much stronger than the probe field. By setting  $\left\langle O \right\rangle=O_s +\delta O\,(O=c_{1,2},\,L_z,\,\phi)$, one can obtain the steady-state values of the corresponding dynamical variables as
\begin{align}
\phi_{s}	&=	\frac{-g_{1}\hbar\left|c_{1s}\right|^{2}+g_{2}\hbar\left|c_{2s}\right|^{2}}{I\omega_{\phi}^{2}},\;L_{zs}=0,\\
c_{1s}	&=	\frac{\epsilon_{1}}{\kappa_1+i\Delta_{1}},\;c_{2s}=\frac{\epsilon_{2}}{\kappa_2+i\Delta_{2}},\\
\Delta_{1}	&=\Delta_{c1}+g_1\phi_{s} \notag \\
&=\left(\Delta_{c1}-\frac{g_1^{2}\hbar \left|c_{1s}\right|^{2}}{I\omega_{\phi}^{2}}\right) +  \label{7}
\frac{g_1 g_2\hbar \left|c_{2s}\right|^{2}}{I\omega_{\phi}^{2}},\\
\Delta_{2}&=\Delta_{c2}-g_{2}\phi_{s}.
\end{align}
in which $\Delta_{1}\,(\Delta_{2})$ represents the effective detuning of cavity $c_1\,(c_2)$ from the driving fields.
One can find that the effective detuning of cavity $c_1$ can be modulated effectively by cavity $c_2$. Besides, this
modulation can be improved if we choose a cavity with higher cavity finesse and stronger driving power, as well as a
mirror with smaller mass and size. Meanwhile, the resonantly driven cavity $c_2$ can also greatly enhance its modulation
effect on cavity $c_1$.

Besides, the equations of the corresponding perturbation terms can be derived as follows,
\begin{align}
\frac{d\delta c_{1}}{dt}	&\!=\!	-(\kappa_1\!+\!i\Delta_{1})\delta c_{1}\!-\!ig_{1}\delta\phi(c_{1s}\!+\!\delta c_{1})\!+\!\epsilon_{p}e^{-i\Omega t},\\
\frac{d\delta c_{2}}{dt}	&\!=\!	-(\kappa_2+i\Delta_{2})\delta c_{2}+ig_{2}\delta\phi(c_{2s}+\delta c_{2}),\\
\frac{d^{2}\delta\phi}{dt^{2}}	&\!=\!	-\gamma_{\phi}\frac{d\delta\phi}{dt}-\omega_{\phi}^{2}\delta\phi \notag \\
&-\frac{\hbar g_{1}}{I}(c_{1s}^{*}\delta c_{1}+\delta c_{1}^{*}c_{1s}+\delta c_{1}^{*}\delta c_{1}) \notag \\
&+\frac{\hbar g_{2}}{I}(c_{2s}^{*}\delta c_{2}+\delta c_{2}^{*}c_{2s}+\delta c_{2}^{*}\delta c_{2}).
\end{align}
The above equations of the perturbation terms can be solved by applying the ansatz, i.e., $\delta O=O_{+}e^{-i\Omega t}+O_{-}e^{i\Omega t}$, then, one can get the solution of $c_{1+}$, which corresponds to the response of the system to
the probe field \cite{agarwal2010electromagnetically,Huang2010Normal}.
\begin{eqnarray}
c_{1+}\!=\!\frac{-i\epsilon_{p}[N_{1}g_{1}^{2}\hbar\!+\!2N_{2}\Delta_{2}g_{2}^{2}\hbar D_1(\Omega) \!+\!ID_2(\Omega)]}{2N_{1}\Delta_{1}g_{1}^{2}\hbar\!+\!2N_{2}\Delta_{2}g_{2}^{2}\hbar D_3(\Omega)\!+\!ID_4(\Omega)},
\end{eqnarray}
with
\begin{align}
D_1(\Omega)&=\frac{\Delta_{1}+\Omega+i\kappa_1}{\Delta_{2}^{2}+(\kappa_2-i\Omega)^{2}},\\
D_2(\Omega)&=(\Delta_{1}+\Omega+i\kappa_1)(\Omega^{2}-\omega_{\phi}^{2}+i\gamma_{\phi}\Omega),\\
D_3(\Omega)&=\frac{\Delta_{1}^{2}+(\kappa_1-i\Omega)^{2}}{\Delta_{2}^{2}+(\kappa_2-i\Omega)^{2}},\\
D_4(\Omega)&=[\Delta_{1}^{2}+(\kappa_1-i\Omega)^{2}](\Omega^{2}-\omega_{\phi}^{2}+i\gamma_{\phi}\Omega).
\end{align}

\begin{figure*}
	\centering
	\includegraphics[width=1\linewidth]{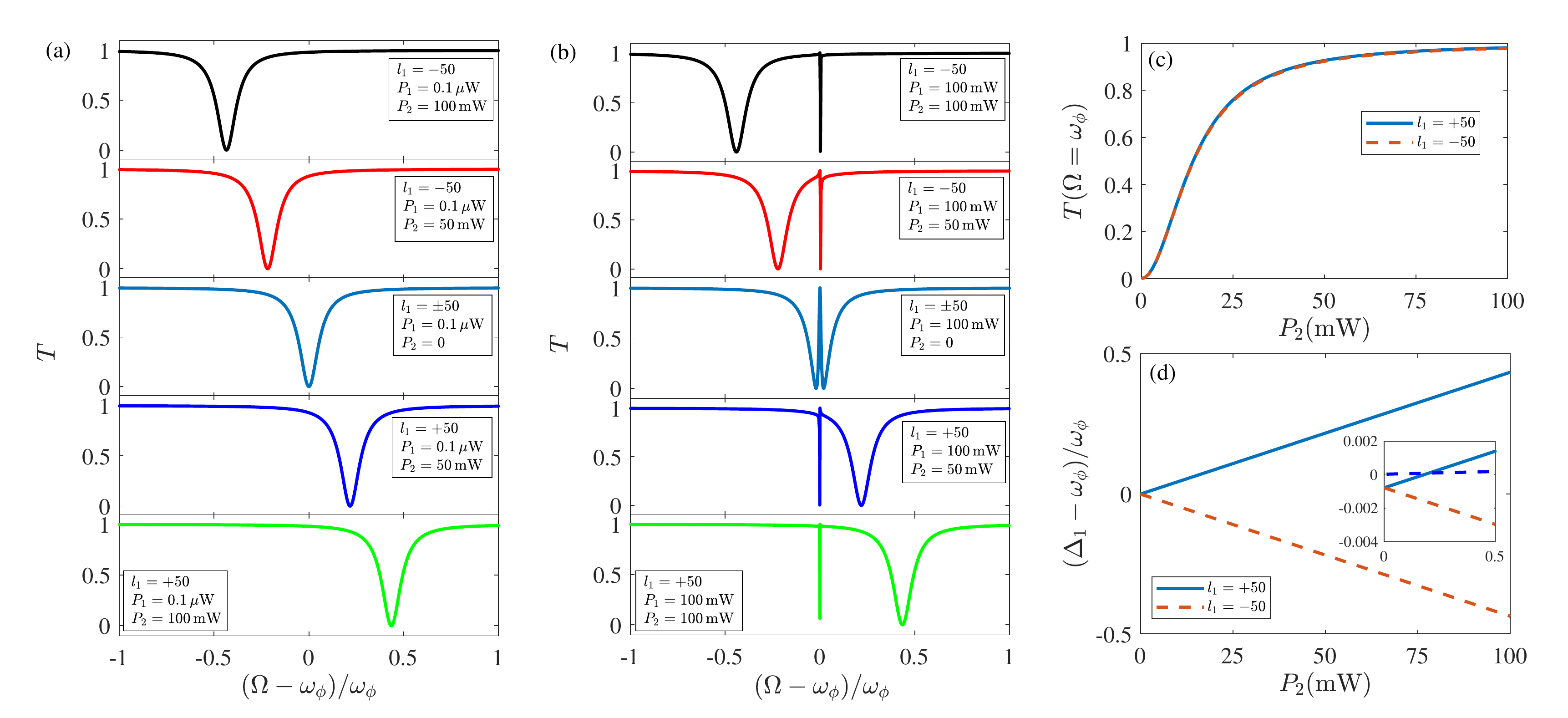}
	\caption{(a), (b) Transmission spectrum of the probe field in the double LG rotational cavity as a function of the normalized detuning $(\Omega-\omega_{\phi})/\omega_{\phi}$ for different driving powers $P_{2}$ of cavity $c_2$ $(P_2=0,\,50\,\textrm{mW},\,100\,\textrm{mW})$, in which
		magenta curve represents $l_{1}=\pm 50$, black curve and red curve represent $l_{1}=-50$, and blue curve and green curve represent $l_{1}=50$. (c) Transmission of the probe field at resonance as a function of the driving power of cavity $c_2$. (d) Normalized cavity detuning as a function of the driving power of cavity $c_2$. Parameters are: (a), (c) $P_{1}=0.1\,\textrm{\ensuremath{\mu}W},\,l_{2}=100$; (b), (d) $P_{1}=100\,\textrm{mW},\,l_{2}=100$; and other parameters are the same as in the main text.}
	\label{fig:f2}
\end{figure*}
Then according to the standard input-output relation \cite{Gardiner2004Quantum}, the transmission of the probe field can be defined as
\cite{agarwal2010electromagnetically,Huang2010Normal}
\begin{equation}
T=\left|\frac{\epsilon_{p}/\sqrt{2\kappa_1}-\sqrt{2\kappa_1}c_{1+}}{\epsilon_{p}/\sqrt{2\kappa_1}}\right|^{2}
=\left|1-2\kappa_1 c_{1+}/\epsilon_{p}\right|^{2}.
\end{equation}
The optical response of the optomechanical system to the probe field can be detected in experiments, for example, the homodyne detection scheme in Ref.~\cite{weis2010optomechanically}.

\section{Transmission spectrum of the probe field in the double LG rotational cavity}\label{3}
In this section, we will study the response of double LG rotational cavity to the probe field, and compare it with the case of the
single LG rotational cavity. The parameters used in our paper are chosen from
Refs.~\cite{bhattacharya2007using,bhattacharya2008entanglement,bhattacharya2008entangling}. For RM,
the radius $R=10\, \mu \textrm{m}$, the mass $M=100\,\textrm{ng}$, the angular frequency
$\omega_{\phi}=2\pi\times10\,\textrm{MHz}$, and the mechanical quality factor $Q_{\phi}=2\times10^{6}$.
For cavity, the cavity length $L=5\,$mm, the cavity finesse $F_{1,2}=5\times10^{4}$, and cavity $c_1$ is driven with red-detuned
driving (i.e., $\Delta_{c1}=\omega_{\phi}$, and the frequency $\omega_{1}$ of driving field is $2\pi c/\lambda_{1}$ with
wavelength $\lambda_{1}=1064\,\textrm{nm}$); meanwhile, cavity $c_2$ is driven resonantly with the effective detuning
$\Delta_{2}=0$. We would point out that cavity $c_1$ with red-detuned driving can exhibit bistability
for a strong enough driving field \cite{mccullen1984mirror,baas2004optical}, however, the parameters
used in our paper are chosen in the area of monostability.

Due to the presence of the \textit{optorotational} coupling, the OMIT phenomenon can be generated in the single LG rotational cavity, and
based on the correlation between the window width and the topological charge, OAM can be
measured in principle \cite{peng2019optomechanically}. But, we would like to point out that only the magnitude of OAM can be measured in
a standard single LG rotational cavity. This is due to the fact that the effective cavity detuning $\Delta_{1}$ of the single LG rotational cavity is only
related to the magnitude of OAM and not to its sign, which can be seen from Eq.~(\ref{7}) of our paper [this equation is simplified as
$\Delta_{1}=\Delta_{c1}-g_{1}^{2}\hbar |c_{1s}|^{2}/(I\omega_{\phi}^{2})$ for the single LG rotational cavity], and Eq.~(4) of
Ref.~\cite{peng2019optomechanically}. For the double LG rotational-cavity system of our scheme, however, the effective detuning of cavity $c_1$ can be
modulated by cavity $c_2$, i.e., the second term of Eq.~(\ref{7}), $g_1 g_2\hbar \left|c_{2s}\right|^{2}/(I\omega_{\phi}^{2})$, which leads to that for different signs
of OAM,  the effective detuning of cavity $c_1$ will be modulated differently, as shown in the following sections.

At first, we study the transmission characteristics of the probe field in the double LG rotational cavity. The transmission
spectrum of the probe field in the double LG rotational cavity is plotted as a function of the normalized detuning
$(\Omega-\omega_{\phi})/\omega_{\phi}$, as shown in Figs.~\ref{fig:f2}(a) and \ref{fig:f2}(b). From the curves of
Fig.~\ref{fig:f2}(a), one can see that for a weak driving field of cavity $c_1$ with power $P_{1}=0.1\,\mu$W, the Lorentzian-shaped
transmission spectrum originally located at the frequency $\Omega \approx\omega_{\phi}$ shifts obviously with the
increase of the driving field of cavity $c_{2}$. What's more, for different signs of OAM carried by the beams in cavity $c_1$, the direction of the spectral shift is
just opposite. Specifically, for the topological charge $l_1=50$, the transmission spectrum shifts to the right, but for
$l_1=-50$, it shifts to the left. Thus, this spectral shift can be served as a fully switchable light switch through adjusting the
power of the driving field, as shown in Fig.~\ref{fig:f2}(c). Besides, one can find from the curves of Fig.~\ref{fig:f2}(b) that for a
stronger driving field of cavity $c_1$ with power $P_{1}=100\,$mW, a symmetrical OMIT window occurs in the frequency close to resonance,
but this symmetry is broken once cavity $c_2$ is introduced, then an asymmetric window similar
to the Fano resonance can be observed. Like the Lorentzian-shaped transmission spectrum [see Fig.~\ref{fig:f2}(a)], the
asymmetric transparency window also has a strong correlation with the OAM carried by the beams in cavity $c_1$. 
This transmission characteristics of the probe field in our system can not be realized with the standard single LG rotational cavity \cite{peng2019optomechanically}.

\begin{figure}
	\centering
	\includegraphics[width=1\linewidth]{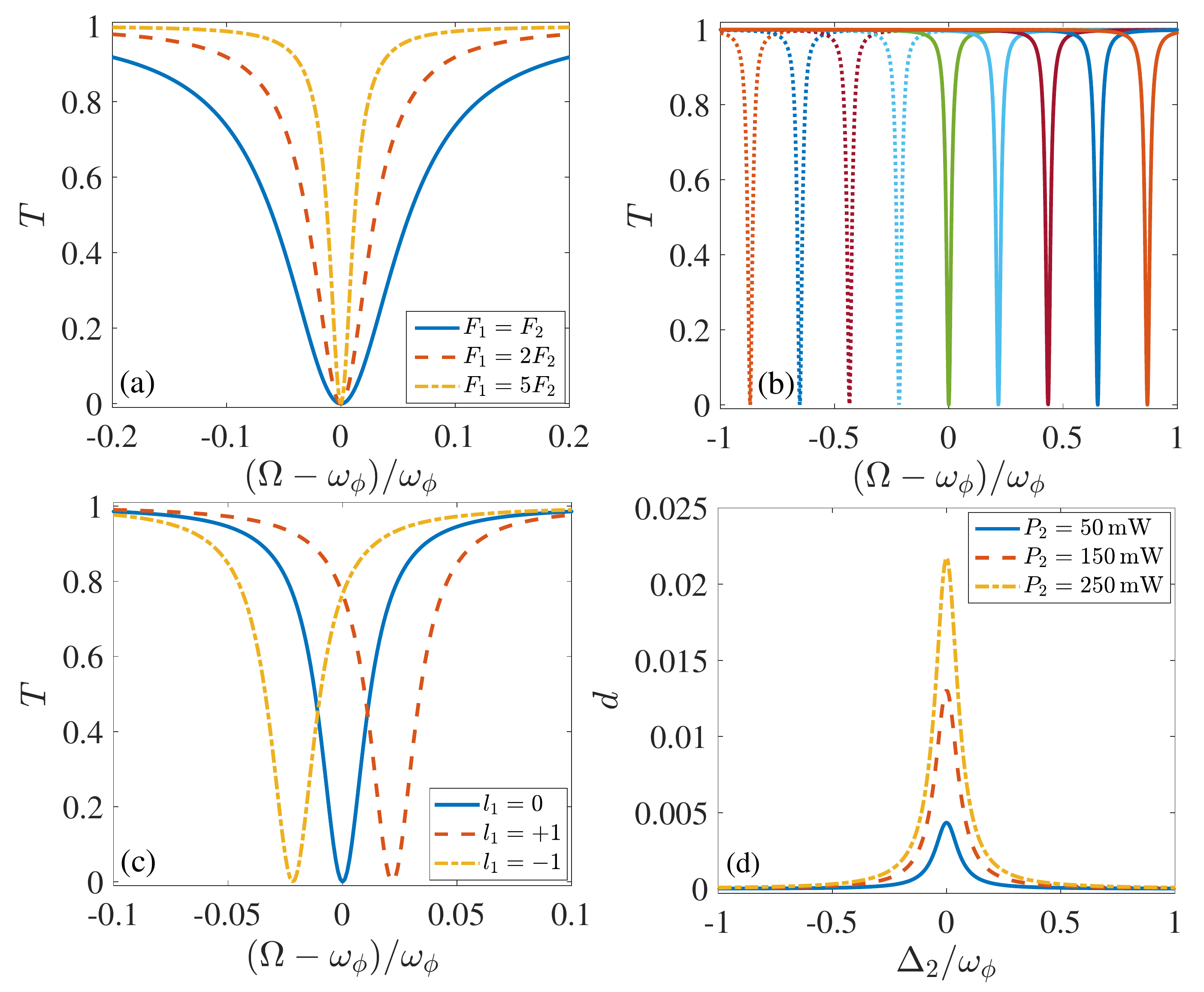}
	\caption{Transmission spectrum of the probe field as a function of the normalized detuning $(\Omega-\omega_{\phi})/\omega_{\phi}$ for (a) different cavity finesse $F_1$,
		(b) different orbital angular momentum (from left to right: $l_{1}=-40,\,-30,\,-20,\,-10,\,0,\,10,\,20,\,30,\,40$), and (c) orbital angular momentum with $l_1=0,$ and $\pm 1$.
		(d) Distance $d$ of the spectral shift caused by the orbital angular momentum change with $|\Delta l_1|=1$ as a function of the normalized detuning $\Delta_{2}/\omega_{\phi}$.
		Parameters are: (a) $P_1=0.1\,\mu$W, $l_1$=0; (b) and (c) $P_1=0.1\,\mu$W, $P_2=250\,$mW, $F_1=5F_2$; (d) $P_1=0.1\,\mu$W. And other parameters are the same as in the main text.}
	\label{fig:f3}
\end{figure}

This correlation can be understood based on the dependence of the effective detuning of cavity $c_1$ on the cavity $c_2$.
As shown in Fig.~\ref{fig:f2}(d), we plot the normalized cavity detuning $(\Delta_{1}-\omega_{\phi})/\omega_{\phi}$ as a
function of the driving power $P_2$. From Fig.~\ref{fig:f2}(d), one can clearly see that the effective detuning of cavity
$c_1$ strongly depends on the driving power of cavity $c_2$, where for different signs of OAM it changes in reverse with the
increase of the driving field. Meanwhile, one can also find from the inset of Fig.~\ref{fig:f2}(d) that when the value of the driving power of cavity $c_2$
is zero, the value of the normalized cavity detuning of cavity $c_1$ isn't zero, which is due to the existence of the \textit{optorotational} coupling of
cavity $c_1$. Furthermore, the curves of Fig.~\ref{fig:f2}(d) show that the value of the normalized cavity detuning is consistent
with the position of the resonance valley of the transmission spectrum. Thus, there is a strong correlation between the spectral shift
and the OAM carried by the beams of cavity $c_1$ in our system. In addition, the linewidth of the transmission spectrum of the probe field in our system
can be decreased by choosing the cavity $c_1$ with a high cavity finesse, as shown in Fig.~\ref{fig:f3}(a), which will contribute to measuring the OAM with a higher
resolution. The above directional spectral shift induced by cavity $c_2$ gives us the inspiration of measuring both the magnitude and sign of OAM simultaneously.

\section{Measurement of the magnitude and sign of orbital angular momentum}\label{4}
In this section, we apply the directional spectral shift induced by cavity $c_2$ to measure the magnitude and sign of OAM.
As shown in Figs.~\ref{fig:f3}(b) and~\ref{fig:f3}(c), the transmission spectrum of the probe field is plotted as a function of the normalized detuning
$(\Omega-\omega_{\phi})/\omega_{\phi}$ for a wide range of OAM.
One can observe from the curves that for a fixed driving power of the system, the Lorentzian-shaped
transmission spectrum can show a significant spectral shift when we increase the magnitude of OAM. Specifically, for
OAM with positive sign, the resonance valley of the transmission spectrum shifts to the right with the increase of the
magnitude of OAM, and for OAM with negative sign, the situation is exactly the opposite. Moreover, one can also
observe in Fig~\ref{fig:f3}(c) that our method can distinguish OAM with $\Delta l_1=\pm 1$ by monitoring the shift of the transmission spectrum. As expected,
based on the shift of the transmission spectrum of the probe field in a double LG rotational-cavity system,
the magnitude and sign of OAM can be simultaneously measured compared with the case of the single LG rotational-cavity system in Ref.~\cite{peng2019optomechanically}.
Utilizing the spectral shift of the transmission spectrum to measure OAM, in addition to the linewidth of the transmission spectrum, the distance of
the spectral shift caused by the change of OAM is also an important index. As shown in Eq.~(\ref{7}), the distance of the spectral shift in our scheme
can be enhanced by the resonantly driven cavity $c_2$ with a strong driving power.
This comes from the fact that the intracavity photon number in cavity $c_2$ can be increased dramatically in this case, which can enhance its modulation
effect on cavity $c_1$. The distance $d$ of the spectral shift caused by the OAM change with $|\Delta l_1|=1$ is plotted as a function of the normalized
detuning $\Delta_{2}/\omega_{\phi}$, as shown in Fig.~\ref{fig:f3}(d).
One can see that the distance of the spectral shift can get its maximum value at the resonance frequency $\Delta_{2}=0$,
and the maximum value can be increased by the driving power $P_2$. Then, with the optimized distance of the spectral shift and its linewidth, our method can distinguish OAM with a higher sensitivity.

\begin{figure}
	\centering
	\includegraphics[width=1\linewidth]{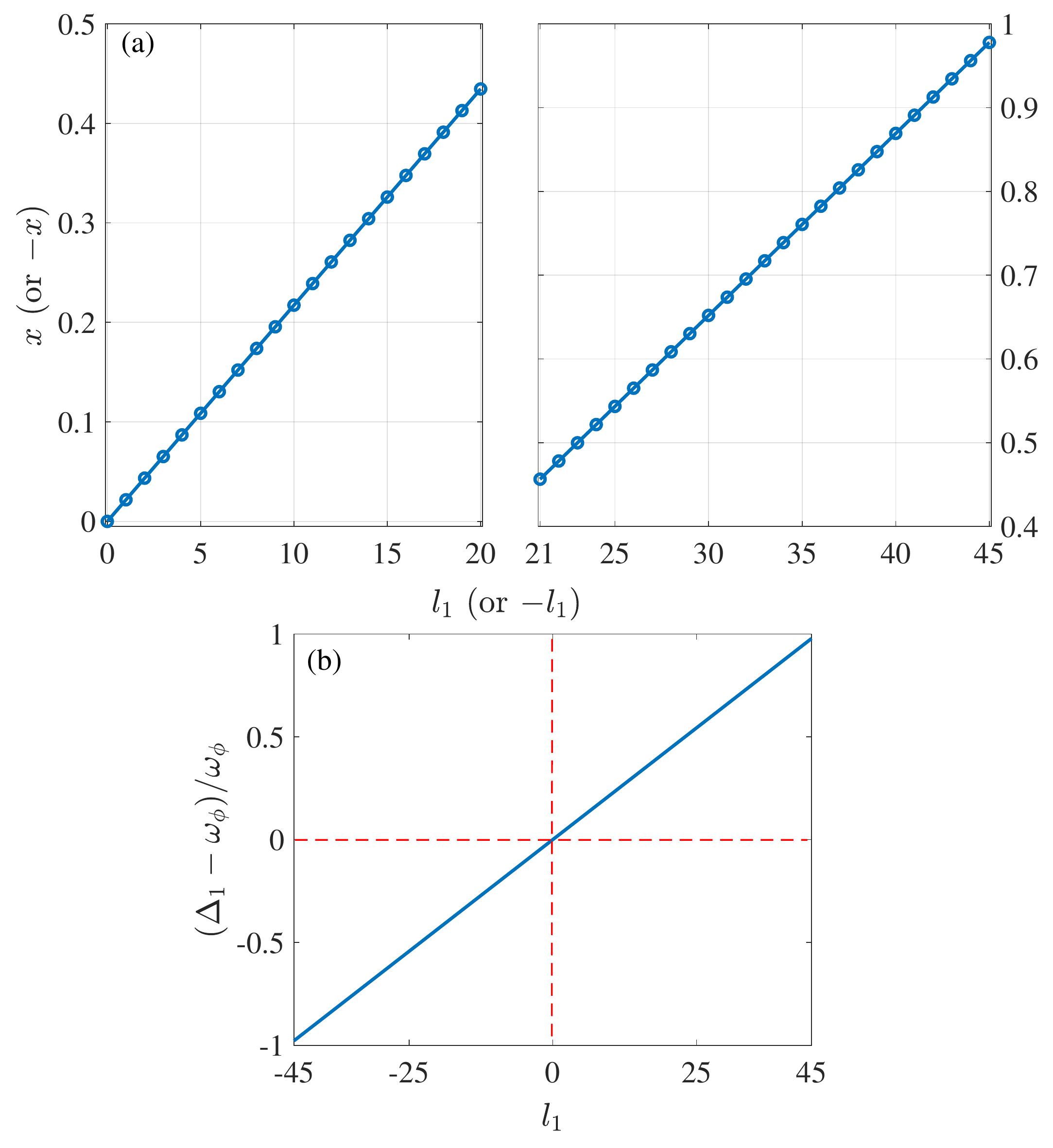}
	\caption{(a) Position of the resonance valley of the transmission spectrum as a function of the orbital angular momentum. (b) Normalized cavity detuning as a function of the orbital angular momentum.
		Parameters are the same as in Fig.~\ref{fig:f3}.}
	\label{fig:f4}
\end{figure}

Thus, based on the correlation between position of the resonance valley and OAM, we can measure the magnitude and sign of OAM
simultaneously. The position of the resonance valley can be determined by the following conditions,
\begin{equation}
\frac{dT}{dx}=0, \quad \frac{d^{2}T}{dx^2}>0,
\end{equation}
with $x=(\Omega-\omega_{\phi})/\omega_{\phi}$ corresponding to the position of the resonance valley.
In order to clearly show this correlation, we plot the position of the resonance valley as a function of the
magnitude and sign of OAM, as shown in Fig.~\ref{fig:f4}(a), where the topological charge value $l_1$ changes in
integer. Figure~\ref{fig:f4}(a) shows that there is an almost linear relationship between the position of the resonance valley and
the magnitude of OAM. Meanwhile, for the positive and negative signs of OAM, the position is basically symmetrical about
the resonance frequency $\Omega=\omega_{\phi}$, which is different from Fig.~5 of Ref.~\cite{peng2019optomechanically}, where only the magnitude of
OAM can be measured. Moreover, with the spectral shift in our system, the measurable topological charge is up to $\pm 45$.
This spectral shift with the magnitude and sign of OAM can be understood with the effective
detuning of cavity $c_1$, i.e.,  Eq.~(\ref{7}). As shown in Fig.~\ref{fig:f4}(b), the normalized cavity detuning of cavity $c_1$
is plotted as a function of the OAM. One can find that there is an almost linear relationship between the value of the normalized cavity detuning and
the magnitude of OAM, and the trend of the normalized cavity detuning with OAM (including its magnitude and sign) is consistent with Fig.~\ref{fig:f4}(a).

Based on the above analysis, by monitoring the shift of the transmission spectrum of the probe field in a double LG rotational-cavity system,
our scheme can measure a wider range of OAM with distinguishable sign compared with the case of the single LG rotational cavity in Ref.~\cite{peng2019optomechanically}.
Meanwhile, the sensitivity of our scheme can be improved by optimizing the distance of the spectral shift and its linewidth,
which is a significant improvement for measuring OAM in optomechanics.

\section{Conclusions}\label{5}
In summary, we have investigated the transmission characteristics of the probe field in the double LG rotational cavity and
showed that the effective cavity detuning in this system depends on the magnitude and sign of OAM simultaneously, which is different
from the case of the single LG rotational cavity. Moreover, we found that the transmission spectrum of the probe field
has a strong correlation with the magnitude and sign of OAM. Specifically, for different magnitudes of OAM, the
transmission spectrum can show an obvious spectral shift, meanwhile, the spectral shift is directional for different signs
of OAM. Thus, based on this feature, we propose a scheme to measure OAM including the magnitude and the sign, in
which the measurable topological charge is up to $\pm 45$. This work solves the shortcoming of the inability to distinguish the sign of OAM in optomechanics.
 
\section*{acknowledgments}
	This work was supported by the National Key Research and Development Program of China (Grants No.~2017YFA0304202 and No.~2017YFA0205700), the National Natural Science Foundation of China (NSFC) (Grants No.~11875231 and No.~11935012), and the Fundamental Research Funds for the Central Universities through Grant No.~2018FZA3005.

\bibliography{references}

\end{document}